\documentstyle[prl,twocolumn,aps]{revtex}

\begin{document}
\tolerance 50000
\draft
\twocolumn[\hsize\textwidth\columnwidth\hsize\csname @twocolumnfalse\endcsname

\title{Magnetic field induced directional localization
in a 2D rectangular lattice}

\author{A. Barelli$^1$, J. Bellissard$^1$ and F. Claro$^{2}$ }

\address{$^{1}${\it Laboratoire de Physique Quantique, UMR5626 associ\'ee au CNRS,
IRSAMC, Universit\'e Paul Sabatier, 118, route de Narbonne, 
F-31062 Toulouse Cedex 4, France}\\
$^{2}${\it Facultad de F\'{\i}sica, Pontificia Universidad
 Cat\'{o}lica de Chile, Vicu\~{n}a Mackenna 4860, Casilla 306,
 Santiago 22, Chile}}

\maketitle

\begin{abstract}
We study the effect of a perpendicular uniform magnetic field 
on the dissipative conductivity of a rectangular lattice with 
anisotropic hopping, $t_x\neq t_y $. We show that the magnetic 
field may enhance dramatically the directional anisotropy
in the conductivity. The effect is a measurable physical 
realization of Aubry's duality in Harper systems. 

\end{abstract}

\vskip2pc]

Localization in aperiodic systems has been at the center of attention for
decades in condensed matter physics. Most work concerned disordered systems.
Twenty years ago, however, Aubry et al. \cite{aubr} predicted that a 1D
tight binding hamiltonian with a quasiperiodic potential exhibits a
metal-insulator transition as the amplitude of the potential becomes larger than
a critical value. The proof rests on a duality property that allows mapping low
into large coupling constants, while corresponding extended wave functions are
transformed into states that are localized. The tight  binding model used by
Aubry leads to the almost Mathieu equation \cite{Simon,gap2}, a special case of
a more general class of quasiperiodic systems for which the duality applies. It
also happens that the almost Mathieu equation arises in the study of the
dynamics of electrons in 2D in the presence of a rectangular lattice and a
perpendicular uniform magnetic field (see \cite{harp} for the case of a square
lattice). In this case Aubry's duality may be interpreted as a rotation by
$\pi/2$ of the lattice, an operation that can be performed easily in a real
sample and thus lends itself to experimental test. The aim of this letter is to
show that due to Aubry's duality, turning on a magnetic field may produce a
dramatic enhancement of the anisotropy already present in the conductivity of a
rectangular potential. 
We denote by $t_x,t_y$ the hopping amplitudes along the $x$ and $y$ axis (with
$t_x < t_y$), and by $\sigma_{xx}, \sigma_{yy}$ the corresponding longitudinal
conductivities. At zero magnetic field the Drude formula 
away from the parabolic edges of a tight binding band yields
$\left( \sigma_{xx}/\sigma_{yy}\right) _o\approx const \left( t_x/t_y\right)^2$
for small $t_x/t_y$. With magnetic field, however, we obtain
within the relaxation time approximation (RTA)

\begin{equation}
\label{ratiocond}
\frac{\sigma _{xx}}{\sigma _{yy}}
  \Big\vert _B
 \approx
  \frac{\gamma}{2}
   \left(
     \frac{\hbar}{\tau}\pi n(\mu)
   \right)^2
    \frac{\sigma _{xx}}{\sigma _{yy}}
 \Big\vert _0   
    \mbox{ , }
\end{equation} 

\noindent for irrational flux per plaquette. Here $n(\mu)$ is the
density of states per unit cell at the Fermi energy $\mu$,
$\tau$ is the dominant scattering time and $\gamma$ is a constant of $O(1)$. 
In deriving this result we assume that the temperature is low enough so that
$kT < \hbar/\tau$, and that the Fermi level is not too close from the edge of a gap larger than or
equal to $O(k_B T,\hbar/\tau )$. It may fail as well in a regime where Mott's variable range hopping dominates. 
The enhanced asymmetry exhibited by Eq. (\ref{ratiocond}) for a large relaxation time
is physically understandable in terms of Aubry's duality: the electronic
eigenstates are extended along the easy direction $y$ leading to a metallic-like
behaviour for $\sigma_{yy}$, whereas they are localized in the direction $x$,
leading to an insulating-like behaviour for $\sigma_{xx}$.  We shall
argue that the predicted enhancement should be observable in a superlattice of
quantum dots.

Consider a tight-binding hamiltonian in the x-y plane. For
convenience we choose the gauge $\vec{A_x}=B(0,x,0)$ in which the coordinate $y$
becomes cyclic, permitting plane wave solutions along this latter axis. The wave
function along $x$ must then obey the almost Mathieu difference equation  \cite{foo1}

\begin{eqnarray}
\label{almostmathieu}
2t_y\cos(2\pi\alpha m -k_yb)\psi(ma)+\nonumber\\
t_x\left[\psi((m+1)a) + \psi((m-1)a)\right]=E\psi(ma)\mbox{ . } 
\end{eqnarray}

\noindent Here the field variable $\alpha=eBab/hc$ gives the number of flux quanta
traversing the unit cell, $E$ is the energy, $a$ and $b$ are the lattice
constants in the $x$ and $y$ direction, respectively, $k_y$ is the wave number
of the free running plane wave along the $y$ axis, and $m$ is an integer
labelling the lattice sites. The conventional Harper model \cite{harp} is
obtained by making the lattice square, with $a=b$ and $t_x = t_y$. We are
interested in the asymmetric case, usually arising from the unit cell being
rectangular, although a square array of elliptical quantum dots, for example,
would also provide the required asymmetry. 

Inspection of Eq.~(\ref{almostmathieu}) shows that in the limit $ t_x/t_y\gg 1
$, the solutions are plane waves, slightly modulated by the quasiperiodic
potential. In the other  extreme $t_x/t_y\ll 1 $ however, the solutions are
localized features a distance $qa$ apart if $\alpha=p/q$ is rational, or a
single  localized feature if this parameter is irrational \cite{locdeloc}. For $
t_x/t_y $ finite one can show that under Fourier transformation
Eq.~(\ref{almostmathieu}) formally turns into itself, with the roles of $t_x$
and $t_y$ exchanged. An extended state obtained for $ t_x/t_y\gg 1 $ is thus
replaced by  a localized state in Fourier transformed space. This property is
known as Aubry's duality \cite{aubr}. Another way of obtaining an exchange of
roles of $t_x$ and $t_y$ is to change the gauge.  To see this assume that $
t_x/t_y\ll 1$ so that the states given by Eq. (1) are localized along the x
axis. If in the original problem one uses the gauge $\vec{A_y}=B(-y,0,0)$
instead of $\vec{A_x}= B(0,x,0)$, the resulting equation is formally identical
to Eq. (1), only that now the wave function describes the dynamics along y, and
$t_x, t_y$ exchange places. Because of this latter fact and our assumption about
the relative size of these parameters the new version of Eq. (1) gives now
extended states that run along the y direction. Thus, while one gauge yields
states that are localized along the hard hopping direction, the other gauge
yields extended states along the easy hopping direction. This means that Aubry's
duality is manifested in a single sample, its two principal axes playing the
r\^ole of the dual states. As we show below, the anisotropy in the conductivity
may reveal this effect in a dramatic way.

\begin{figure}
\vbox to 90bp {\vfil
\hbox{\hbox to 120bp 
{\includegraphics{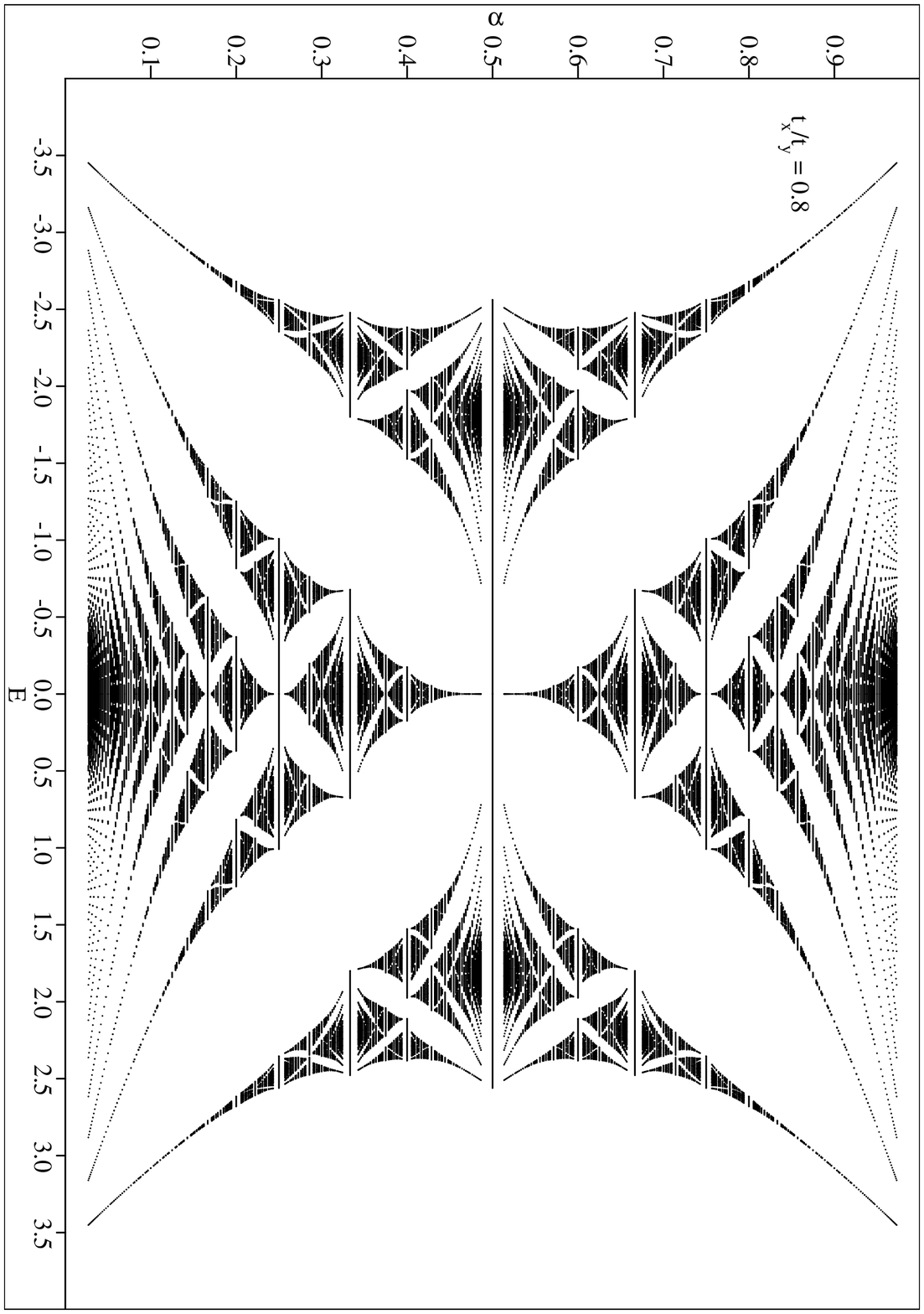}\hfil}
\hbox to 120bp 
{\includegraphics{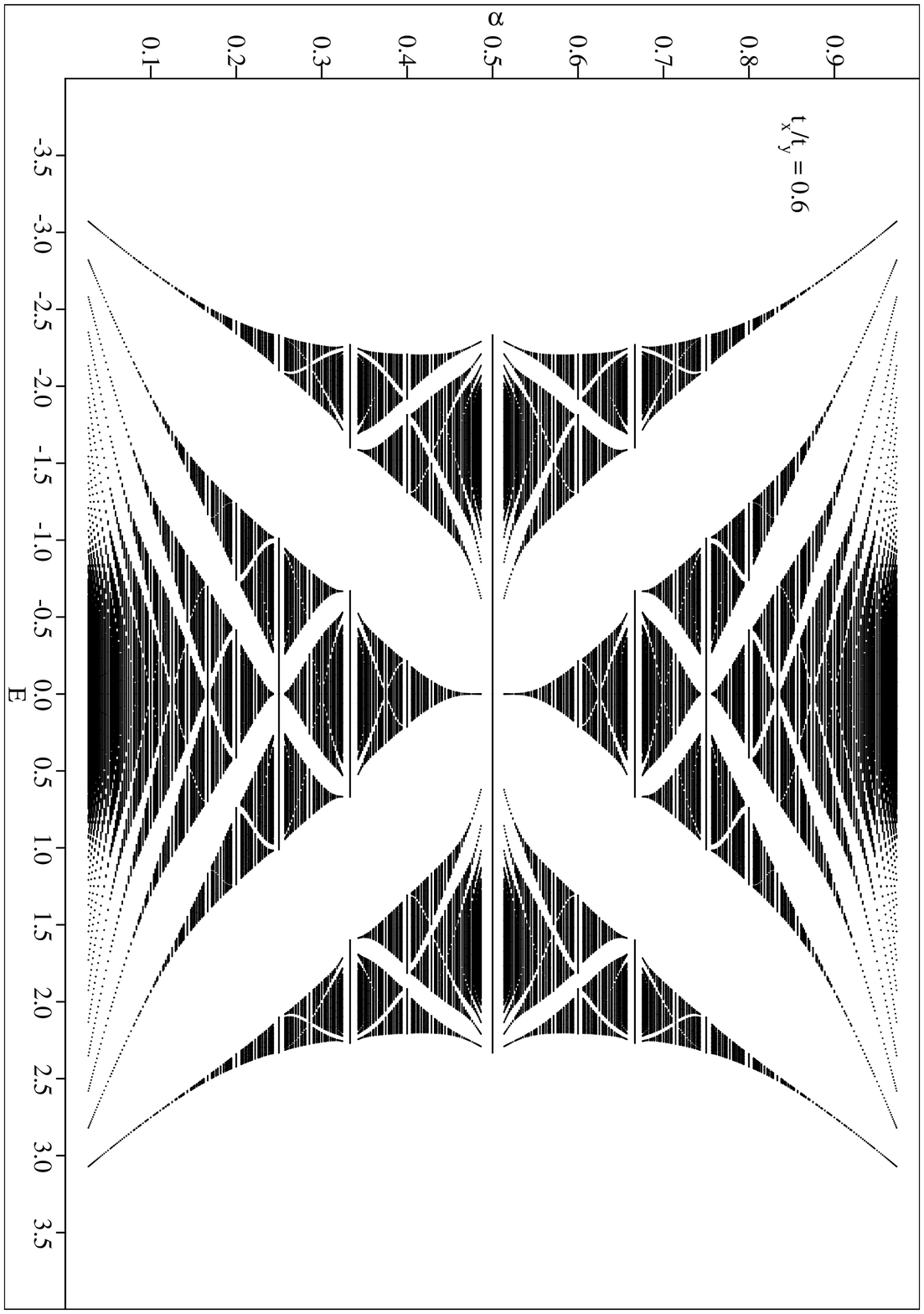}
\hfil}}}
\vbox to 90bp {\vfil
\hbox{\hbox to 120bp 
{\includegraphics{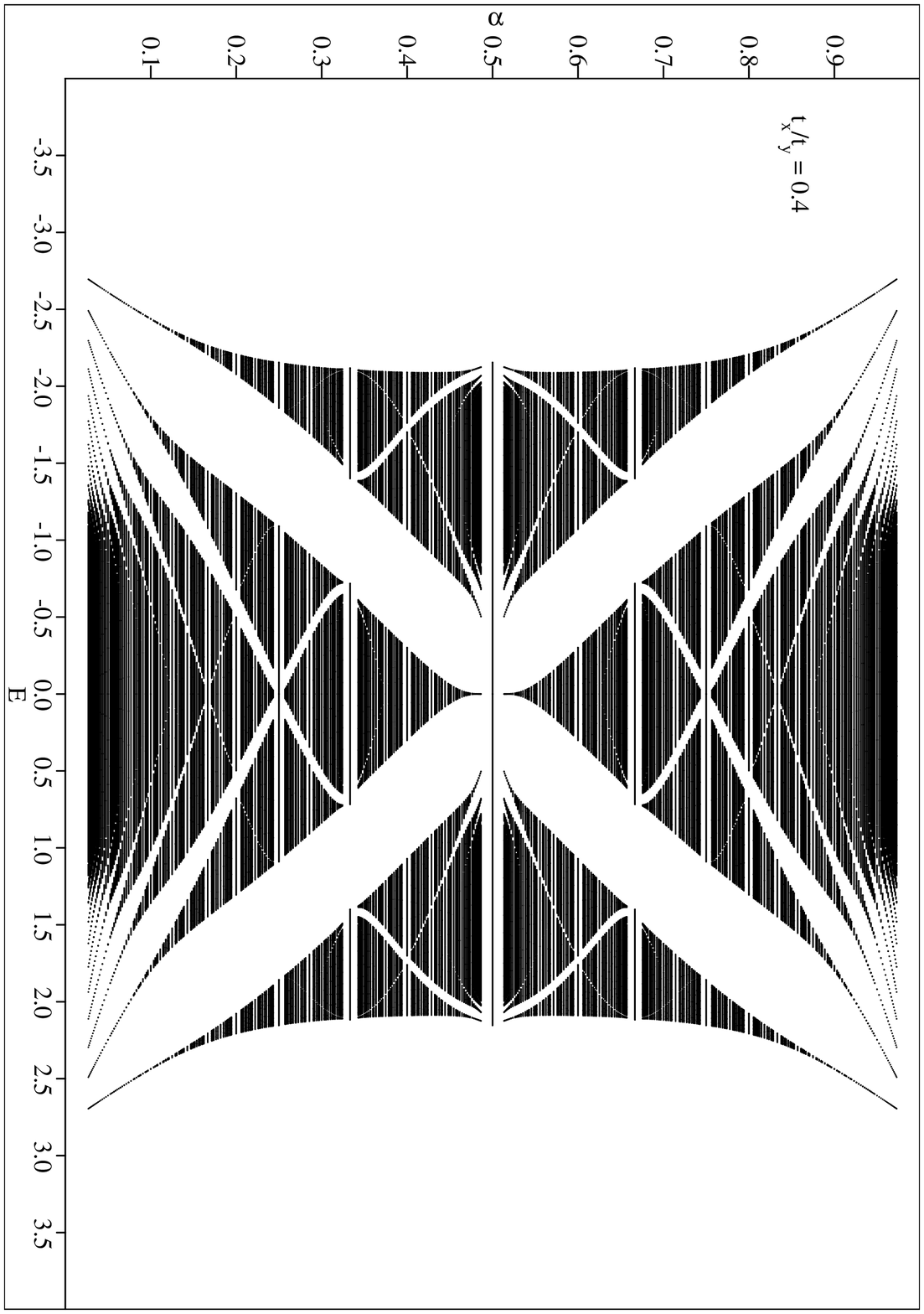}\hfil}
\hbox to 120bp 
{\includegraphics{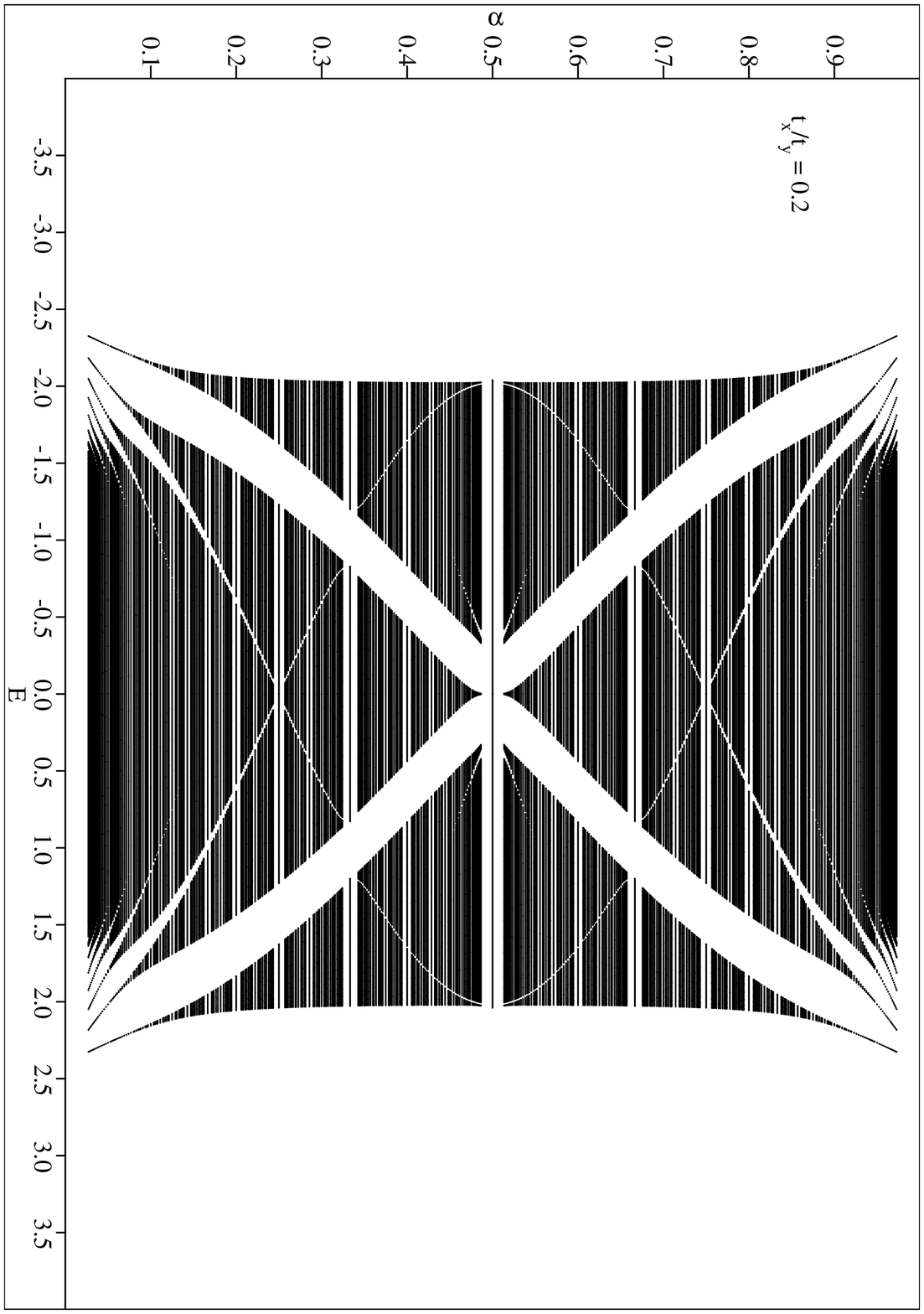}
\hfil}}}
\caption{\label{butterfly}Spectra of the Harper problem on a rectangular lattice
for $ t_x/t_y=0.8, 0.6, 0.4 $ and $ 0.2 $ obtained for  rational values of $
\alpha =p/q $ with $ q\leq 37 $.} 
\end{figure}

The spectrum of Eq.~(\ref{almostmathieu}) for irrational values of the field
parameter is a Cantor set with a hierarchy of gaps that become extremely small
\cite{Simon,gap2,locdeloc}. Fig.~\ref{butterfly} shows the spectrum for
different values of $ t_x/t_y $. Note that as this ratio decreases from 0.8 to
0.2 the spectral gaps become more and more invisible due to
their diminishing size.
Each gap is uniquely labelled by an integer $s$ \cite{gap1} taking values
between $-(q-1)/2$ and $(q-1)/2$ if $\alpha =p/q$, and all values if $\alpha$ is
irrational. It has been shown that the gap width behaves like \cite{gap2}

\begin{equation}
\label{gapsize}
\Delta_s
 \sim
  (t_x/t_y) ^{|s|}t_y
\mbox{ , }
\end{equation}

\noindent for small $ t_x/t_y $. For a rational $\alpha=p/q$, the spectrum has
exactly $q$ subbands that do not overlap so that up to $ q-1 $ gaps may appear.
These are actually all open except for the one at $E=0$ for q
even\cite{wann,gap2}. For irrational $\alpha$ the spectrum can be well
approximated by the rational approximants $p_n/q_n$ \cite{Elliott} obtained from
truncating the continuous fraction expansion of $\alpha$ at its $n$-th step
\cite{Hardy}. The gap labels are then stable through this approximation
\cite{gap1,Elliott}.

\noindent In real samples the presence of scattering limits the experimental access to
such a fine structure \cite{enss}. There is thermal broadening of size $k_BT$,
and also an energy width $\hbar /\tau$ associated with other sources of
scattering, with $\tau$ a characteristic relaxation time. All measurable
quantities will be rather insensitive to gaps smaller than $\delta =
\max{(k_BT,\hbar /\tau)}$. Moreover, only energies within an interval of size
$\delta$ from the Fermi energy $\mu$ will matter for the electronic transport.
Therefore, if $\alpha$ is irrational, it is sufficient to replace it by its 
best rational approximant $p/q$
such that the gaps closest to $\mu$ have width not smaller than $\delta$. The
error introduced in transport properties by this substitution is of the order of
the hopping probability between sites at distance $qa$ apart, which for $t_x$
small is bounded to be also very small. In our calculation of  the conductivity
we shall adopt this simplifying criterion. 

More precisely, for each integer $n$ let
$\Delta^{(n)}$ be the minimal width of the closest gaps to $\mu$ corresponding
to the rational approximant $p_n/q_n$, and let $s_n$ be the corresponding 
gap label. Then one chooses the
largest value $N$ of $n$ such that 

\begin{equation}
\label{choiceofq}
\Delta^{(N+1)} 
  < \delta = \max{(k_B T, \frac{\hbar}{\tau})}
  \leq \Delta^{(N)}
\mbox{ , }
\end{equation}

\noindent thus fixing the values of $q=q_N$ and $s=s_N$. In particular, as the
temperature $T$ decreases, one expects $\tau$ to increase, so that the value of
$N$ increases as well and with it, those of $q_N$ and $s_N$.

A convenient form of Eq.~(\ref{almostmathieu}) for the rational case is
\begin{eqnarray}
\label{rationalHarper}
2t_y\cos(2\pi m\frac{p}{q}-k_yb)\phi_\ell(m)+ 
t_x\left[{\rm e}^{ik_xa}\phi_\ell(m+1)+\right.\nonumber\\
\left.{\rm e}^{-ik_xa}\phi_\ell(m-1)\right]
=E_\ell(k_x,k_y)\phi_\ell(m)\mbox { , }
\end{eqnarray}

\noindent where $m$ is an integer, $\phi_\ell(m)$ is periodic of period $q$, and $k_x$ is
a phase. For each point $\vec{k}=(k_x,k_y)$ in phase space, with $0\leq
k_xa,k_yb\leq 2\pi $, there are $q$ eigenvalues which, as $\vec{k}$  covers its
range, make up the $q$ subbands labelled by the integers $\ell = 1, 2, ..., q$.
All eigenvalues are contained in the energy interval $\vert
E_\ell\vert<2(t_x+t_y)=W/2$, with $W$ the width of the original zero-field band.
We assume the Fermi energy $\mu$ to lie somewhere within this range.

In the infinite volume limit and in the relaxation time approximation, 
the longitudinal conductivity for our $q$ subbands system, is
given by Kubo's formula \cite{foo2} 
\begin{eqnarray}
\label{kubo}
\sigma_{ii}=\frac{2e^2\tau}{q\hbar^2}\sum_{\ell=1}^{q} 
\int \frac{d^2k}{4\pi^2}\left\vert\frac{\partial E_{\ell}} 
{\partial k_i}\right\vert ^2  
\delta (E_{\ell}-\mu)+ \nonumber\\  
\frac{4e^2}{q\hbar^2}
{\rm Re}{\sum_{\stackrel{\ell\ne\ell '=1}{E_{\ell '}<\mu <E_{\ell }}}^{q}} 
\int\frac{d^2k}{4\pi^2}
\frac{\left\vert\langle\phi_{\ell'}\vert\partial \phi_{\ell}/\partial k_i
\rangle\right\vert ^2\left(E_{\ell}-E_{\ell'}\right)}
{1/\tau-\imath\left(E_{\ell}-E_{\ell'}\right)/\hbar}\mbox{ , }
\end{eqnarray}

\noindent where $ E_{\ell } $ and $ E_{\ell '} $ depend upon the phase space coordinates
$(k_x,k_y) $ and the integrals are taken over all of phase space. Following
condition (\ref{choiceofq}), in Eq.~(\ref{kubo}) we have replaced the Fermi 
distribution by a step function. The first
term in this equation is the intraband term, while the second includes all
interband contributions. In the absence of a magnetic field there is just one
band in our model and only the first term is relevant. In what follows we shall
assume $t_x<t_y $.

A careful analysis of Eq.~(\ref{kubo}) shows that the conductivity depends
strongly on the position of the Fermi energy with respect to subband edges. We
distinguish two cases.

\noindent {\bf 1) $\mu$ far from the subband edges: } assume first that the
Fermi energy is at a distance larger than $ {\rm O}\left(\Delta ^2/t_y\right) $
from the nearest gap. The intraband term can be computed solving the eigenvalue
equation 

\begin{equation}
\label{chambers}
P_q (E) = 2 (-)^{q+1}\left( t_x^q\cos{qak_x} + t_y^q \cos{qbk_y} \right)
\mbox{ , }
\end{equation}

\noindent for the appropriate subband $\ell$. In this expression $P_q$ is 
the Chambers polynomial
associated with Eq.~(\ref{rationalHarper}) \cite{cham,gap2}. Due to condition
(\ref{choiceofq}), the interband contribution in Eq.~(\ref{kubo}) can be
expanded in $ 1/\tau $ since $ \hbar /\tau < \vert E_{\ell }-E_{\ell '}\vert $.
Keeping just the first terms in this expansion, one gets to lowest
order in $\lambda=t_x/t_y$ 

\begin{eqnarray}
\label{sigmaXX}
 \sigma_{xx}=
  \left(
   \frac{e}{\pi\hbar}
   \right)^2 
    \frac{2 r \tau}{n(\mu)}
     \frac{\lambda^{2q}}{(1-\xi^2)}+
  \gamma
   \frac{e^2 r n(\mu )}{\tau}
    \lambda^{2}\\
\label{sigmaYY}
 \sigma_{yy}=
  \left(
     \frac{2e}{\pi\hbar}
  \right)^2
   \frac{\tau}{r n(\mu)}
\mbox{ . }
\end{eqnarray}

\noindent Here, as mentioned earlier, $ n(\mu) $ is the density of states at the Fermi energy and $
\gamma$ is a numerical factor of order $ 1 $, while $r=a/b$ and $\xi=P_q(\mu)/2t_y^q$. The
first term in these expressions is the Drude conductivity coming
from the intraband transitions in Eq.~(\ref{kubo}). That the ratio 
$\left( \sigma_{xx}/\sigma _{yy} \right)_{\rm Drude}$ is of
${\rm O}(\lambda^{2q})$ is apparent from the energy derivative in Eq.~(\ref{kubo})
and the fact that the Chambers polynomial depends
on the phases through the constant term in the right hand side 
of Eq.~(\ref{chambers}), only. Also, that the lowest order contribution
to the interband transitions is ${\rm O}(\lambda^{2})$ follows
from the fact that in a perturbative expansion in terms of $\lambda$,
the zeroth order term in $\phi_{\ell}$ does not depend on the
phases.  

Assuming $q_N>1$ and having in mind that at sufficiently low temperatures and 
within the rational approximation ansatz $|s_N|\leq
(q_N-1)/2 $ and $ \hbar /\tau t_y  \sim \Delta^{(N)}/ t_y \sim \lambda^{|s_N|}\geq
\lambda^{(q_N-1)/2} $, one finds the ratio between the Drude and interband
contributions to the conductivity along the x direction to be negligible. Ignoring the
first term in (\ref{sigmaXX}) we then have

\begin{eqnarray}
\label{conductivityratio}
\frac{\sigma _{xx}}{\sigma _{yy}}
 \approx 
  \gamma
   \left(
     \frac{\hbar}{\tau} \frac{\pi r n(\mu )}{2} \lambda
   \right)^2
   \mbox{ . }
\end{eqnarray} 

\noindent
This important result states that as long as $\mu$ is not too close to
a subband edge of the proper rational approximant, the conductivity
ratio vanishes quadratically in the inverse relaxation time for small $\lambda$.
Applying Eq. (6) to the original tight binding band to obtain the corresponding 
conductivity ratio in the absence of a magnetic
field near the band center one then arrives at the form given by Eq. (1).

\noindent {\bf 2) $\mu$ near a subband edge: } when the Fermi energy $\mu$ is
very near a subband edge the situation changes significantly. The density of
states has a logarithmic singularity in that neighborhood, and $ n(\mu )\propto
(t_y/t_x)^{q_n/2} $ in a region of order $ \Delta ^2/t_y $ from the edges. As a
result, Eqs. (4)-(5) are no longer correct. The expressions for the Drude
contribution are valid in the region $0<|\xi|<1-\lambda^{q_N}$ only. In the
range $1-\lambda^{q_N} <|\xi|\leq 1+\lambda^{q_N}$ or if the Fermi energy lies
in a gap, $\sigma_{xx}$ may become as large as $ \sigma_{yy} $. A detailed
discussion of this regime will be published elsewhere \cite{bbc}.  These
situations may not be relevant for experiments however, because they only occur
in a very small region of the spectrum. Indeed, when $ t_x<t_y $, the
probability of having $ \mu $ within $ {\rm O}\left(\Delta ^2/t_y\right) $ from
the subband edge behaves like $ (t_x/t_y)^2 $. In addition, for $q_N$ large the
total length of the subbands is given by $ 4\vert t_y-t_x\vert $
\cite{aubr,thou,last}. Thus, choosing the Fermi energy $ \mu $ at random in the
energy interval $ \left[ -W/2,W/2\right] $ gives $ \mu $ in one subband with
relative probability

\begin{equation}
\label{spectralsize}
\frac{\vert t_y-t_x\vert}{t_x+t_y}
 \sim 1-2\frac{t_x}{t_y}
\mbox{ . } 
\end{equation} 

\noindent Thus, the probability of having $ \mu $ in a gap or close to a gap edge is
proportional to $ t_x/t_y $ which is small \cite{bbc}. Moreover, it is likely
that very small fluctuations in the magnetic field will tend to wash away the
effects of $ \mu $ lying in the anomalous regions close to a subband edge or in
a gap. 

In summary, we have shown that in the presence of a perpendicular magnetic field
the field free conductivity asymmetry of a rectangular lattice may be dramatically 
enhanced in a pure sample. This is a physical realization of Aubry's duality
that may be understood in the following way. The magnetic field affects the
hopping of the tight binding electrons through the effective potential of period
$q$ times the appropriate lattice constant. The site energies are then no longer
the same within and the electron has to tunnel a distance equal to this period,
through a potential mismatch that scales with $\lambda^{-1}$ or $\lambda$,
depending on whether transport is along the $x$ or $y$ directions. For an irational
field the tunneling can only be made possible through scattering events,
which are less likely as the relaxation time diverges. Evidence for this behaviour has also been found in
the spreading of a wave packet in the presence of a weak modulation potential in
two dimensions. In the regime in which the single band approximation holds, such
spreading is entirely directional, occuring along the direction of largest
hopping amplitude only \cite{KK}. Our results also shows that the suppression of
the conductivity along one of the principal axes by an external field is
controlled by the ratio of the gap size to the zero-field band width
$(\hbar/\tau)(1/W)\sim (t_x/t_y)^{|s_N|}$.  With present day rectangular arrays
of antidots this quantity is of order one tenth. Thus the effect should
already be visible. It is likely that, as the miniaturization of mesoscopic
technology progresses, this quantity can be made even smaller so that the
magnetic field induced localization may be more easily observed.

\acknowledgements
We would like to thank the ECOS-CONICYT program (grant C94E07) for making this
work possible. One of us (F.C.) also acknowledges partial
support by FONDECYT, Grants 1960417 and 1990425.


\begin{references}{}
\bibitem{foo1} We assume $t_x, t_y >0$, which can always be achieved by means of
a unitary transformation.
\end{references}
\end{document}